\documentclass[twocolumn,letterpaper]{IEEEAerospaceCLS}  

\usepackage[]{graphicx}    
\usepackage[colorlinks=true, urlcolor=blue, linkcolor=black, citecolor=black]{hyperref} 
\newcommand{\ignore}[1]{}  
\usepackage{amsmath,amssymb,amsfonts,amsthm}
\usepackage{balance}
\usepackage{cite}
\usepackage[inkscapelatex=false]{svg}

\begin{document}
\title{Human as an Actuator Dynamic Model Identification}

\author{%
Harrison M. Bonner\\ 
Department of ECE\\
Auburn University\\
Auburn, AL 36849\\
hmb0101@auburn.edu
\and 
Matthew R. Kirchner\\
Department of ECE\\
Auburn University\\
Auburn, AL 36849\\
kirchner@auburn.edu
\thanks{\footnotesize 979-8-3315-7360-7/26/$\$31.00$ \copyright2026 IEEE}              
}

\maketitle

\thispagestyle{plain}
\pagestyle{plain}

\maketitle

\thispagestyle{plain}
\pagestyle{plain}

\begin{abstract}
This paper presents a method for estimating parameters that form a general model for human pilot response for specific tasks. The human model is essential for the dynamic analysis of piloted vehicles. Data are generated on a simulator with multiple trials being incorporated to find the single model that best describes the data. The model is found entirely in the time domain by constructing a constrained optimization problem. This optimization problem implicitly represents the state of the underlying system, making it robust to natural variation in human responses. It is demonstrated by estimating the human response model for a position control task with a quadcopter drone.
\end{abstract}

\tableofcontents

\section{Introduction}

When a human interacts with physical vehicle, e.g. an aircraft or space vehicle, the pilot influences the dynamics of the system. This effectively adds a form of actuator dynamics as the pilot cannot react instantaneously to feedback or other environmental stimuli. A motivating example is that of computing an ``H-V" diagram for a helicopter directly from the dynamics of the vehicle \cite{kirchner2020reachability}. In this case, the H-V diagram alerts the pilot to safe and unsafe flight regimes: An engine failure in an unsafe zone would make autorotation, and hence a safe landing, impossible. If under the assumption a pilot is tasked with performing an emergency autorotation, then the reaction times and response the pilot is capable of executing directly impact the size of the unsafe region. This makes characterizing the human pilot response as a form of actuator critical in calculating these regions.

Early work in the identification of dynamics of a human pilot is found in \cite{tustin1947nature} \cite{mcruer1974mathematical}. These are limited to low-order, linear, time invariant models with a single-input and a single-output. Typically, a frequency space approach is proposed. A review of human pilot models, including a discussion of the history of frequency domain models and modern time domain models is found in\cite{roncolini2023virtual}.

We present a method for estimating the human dynamic model. Unlike previous works in this area, we do not attempt to estimate in the frequency space and instead solve in the time domain using a form of trajectory optimization. This allows for multiple data collects to be used jointly to form a single, common dynamic model.

Critical to the presented method is the incorporation of vehicle dynamics when estimating the pilot model parameters. This contrasts with previous work \cite{jirgl2015dynamic}, where the model is fitted with the pilot `joystick' commands as the output. We seek not to find a pilot model that \textit{outputs} the same commands, rather we seek a pilot model that achieves the same \textit{vehicle response}. If the goal for the pilot, as was the case in \cite{jirgl2015dynamic}, is to achieve a desired aircraft altitude, then the pilot model must be evaluated with respect to the altitude tracking task. By incorporating the vehicle model into the approach, we effectively add a regularization to the model estimation. An example showing the distinction between approaches is given in Figure \ref{counter example}.

Data for use in demonstrating the proposed method is collected by use of a flight simulator. This allows for the creation of scenarios that reflect compound flight maneuvers, for example executing a coordinated turn, that mimic actual flight operations. Finally, the proposed method can easily be generalized to estimate multi-input, multi-output and/or nonlinear models pilot models to be applicable to broad range of flight operations.

\begin{figure}\label{fig:hv}
\centering
\includegraphics[width=\linewidth]{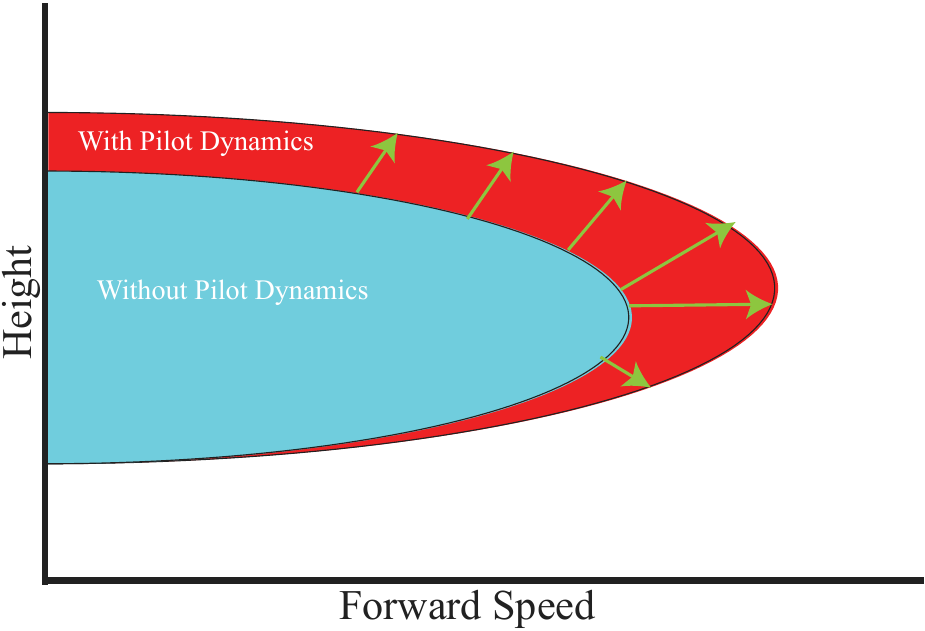}\\
\caption{An illustration of a height-velocity diagram used by helicopter pilots. This shows the unsafe flight regimes as shaded areas. The dynamics of the human pilot must be appended to the system to accurately calculate these diagrams.}
\end{figure}


\section{Problem Formulation}

\begin{figure}\label{counter example}
\centering
\includegraphics[width=\linewidth]{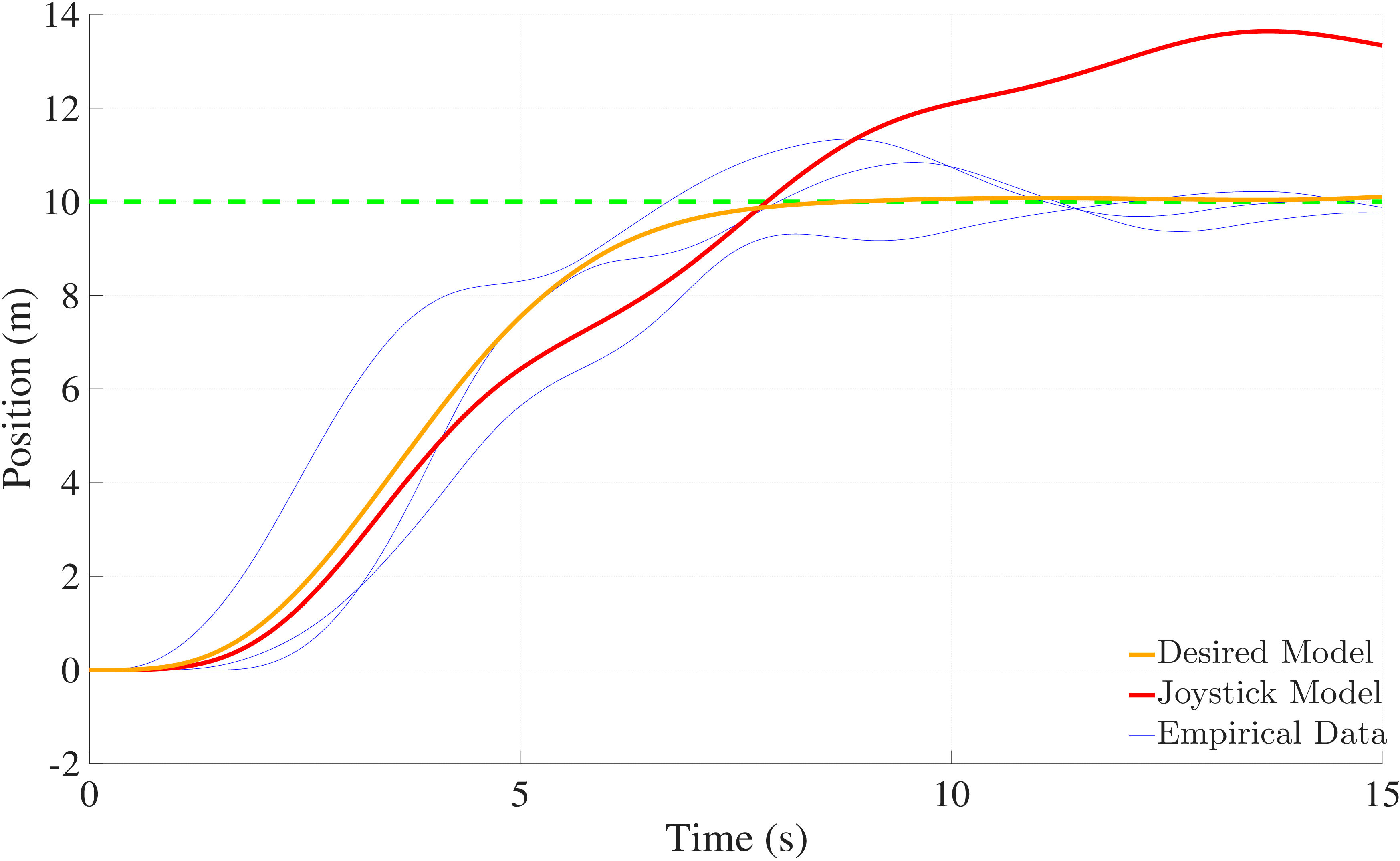}
\caption{An example where the pilot model was estimated using the pilot commands directly. The resulting model applied to the system is shown in red. We can see in the figure that this fails to converge to the goal state. A pilot model fitted using the vehicle dynamics as a regularizer is shown in yellow and is observed converging to the goal position.}
\end{figure}

\begin{figure}\label{fig:flow}
\centering
\includegraphics[width=\linewidth]{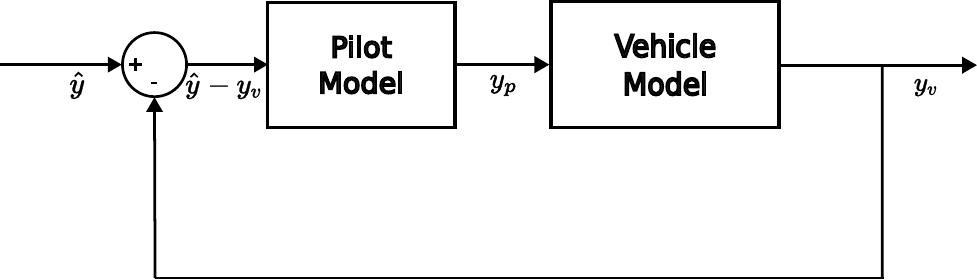}\\
\caption{The graphical depiction of the pilot model and how it forms a joint system with the vehicle.}
\end{figure}

We assume the vehicle of interest has model of the form
\begin{equation}
\begin{cases}
\dot{x}_{v}=f_{v}\left(x_{v},u_{v}\right),\\
y_{v}=h_{v}\left(x_{v}\right),
\end{cases}\label{eq: vehicle dynamics}
\end{equation}
where $x_{v}\in\mathbb{R}^{n_{v}}$ is the state vector and $u_{v}\in\mathbb{R}^{m_{v}}$
is the control input of the vehicle. The variable $y_{v}\in\mathbb{R}^{\ell_{v}}$
is the controlled output. We assume the vehicle is being controlled
by a human pilot that can be modeled as the following dynamic system
\begin{equation}
\begin{cases}
\dot{x}_{p}=f_{p}\left(x_{p},u_{p};\theta\right),\\
y_{p}=h_{p}\left(x_{p};\theta\right),
\end{cases}\label{eq:pilot dynamics}
\end{equation}
where $x_p\in\mathbb{R}^{n_{p}}$ is the state vector and $u_{p}\in\mathbb{R}^{m_{p}}$
is the input of the pilot system. The dynamics of $\left(\ref{eq:pilot dynamics}\right)$
are defined by a vector of parameters $\theta$, and we assume these
parameters remain constant. The pilot is tasked with driving the vehicle's
controlled output, $y_{v}$ to some desired value, $\hat{y}$, i.e.
$y_{v}\left(t\right)\rightarrow\hat{y}$ in finite time. This model structure is shown in Figure \ref{fig:flow}. 

Denote by $x$ the joint state
\[
x:=\left[\begin{array}{c}
x_{v}\\
x_{p}
\end{array}\right].
\]
Then the joint system has dynamics
\begin{align}
\dot{x}:=\left[\begin{array}{c}
\dot{x}_{v}\\
\dot{x}_{p}
\end{array}\right] & =\left[\begin{array}{c}
f_{v}\left(x_{v},y_{p}\right)\\
f_{p}\left(x_{p},\hat{y}-y_{v};\theta\right)
\end{array}\right]\nonumber \\
 & =\left[\begin{array}{c}
f_{v}\left(x_{v},h_{p}\left(x_{p};\theta\right)\right)\\
f_{p}\left(x_{p},\hat{y}-h_{v}\left(x_{v}\right);\theta\right)
\end{array}\right]\\
 & =f\left(x;\hat{y},\theta\right),\nonumber 
\end{align}
where we denote by $f$ the dynamics of the joint system. 

\section{Model Identification With Optimization}

A flight simulator is used to collect data for pilot model identification.
A marker is placed in the environment representing $\hat{y}$, and
the pilot uses this visualization as feedback to maneuver the vehicle
in the simulator to the marker. An example of process is shown in Figure 3. The pilot's joystick command as well
as the vehicle output, $y_{v}$ are recorded by sampling at a fixed
rate. We take $N+1$ total samples over a finite horizon, $\left[0,T\right]$,
on the time grid
\[
\pi^{N}:=\left\{ t_{k}\,|\,k=0,\ldots,N\right\} .
\]
The samples recorded at time $t_{k}$ are denoted as $y_{v}\left(t_{k}\right)$$.$
We do not have direct observation of the system state and follow an
\emph{implicit }formulation for the state vector $x\left(t_{k}\right).$
The measured outputs are assumed corrupted by noise with the following
observation model
\[
y_{v}\left(t_{k}\right)=h_{v}\left(x_{v}\left(t_{k}\right),h_{p}\left(x_{p}\left(t_{k}\right);\theta\right)\right)+\epsilon,
\]
where the noise, $\epsilon$, is assumed i.i.d. Gaussian. We denote
by $X$ as an array formed with each column as state vectors
\[
X:=\left[x\left(t_{0}\right),x\left(t_{1}\right),\ldots,x\left(t_{N}\right)\right],
\]
denote by $\delta Y$ as an array formed with each column as measured
outputs
\[
\delta Y:=\left[\hat{y}-y_{v}\left(t_{0}\right),\ldots,\hat{y}-y_{v}\left(t_{N}\right)\right],
\]
and denote by $H$ as an array formed with each column as 
\begin{align*}
H & :=\Bigg[\hat{y}-h_{v}\left(x_{v}\left(t_{0}\right),h_{p}\left(x_{p}\left(t_{0}\right);\theta\right)\right),\ldots\\
 & \ldots,\hat{y}-h_{v}\left(x_{v}\left(t_{N}\right),h_{p}\left(x_{p}\left(t_{N}\right);\theta\right)\right)\Bigg].
\end{align*}
Lastly denote by $F\left(X;\hat{y},\theta\right)$ as the concatenated
vector:
\[
F\left(X;\hat{y},\theta\right):=\left[\begin{array}{c}
f\left(x\left(t_{0}\right);\hat{y},\theta\right)\\
\vdots\\
f\left(x\left(t_{N}\right);\hat{y},\theta\right)
\end{array}\right].
\]
Since 
\[
\dot{x}\left(t_{k}\right)=f\left(x\left(t_{k}\right);\hat{y},\theta\right)
\]
must hold for each time sample $t_{k}\in\pi^{N}$, we introduce a discrete approximation for $\dot{x}$. Denote by $\mathcal{D}$ the differentiation matrix defined such that 
\[
\left[\begin{array}{c}
\dot{x}\left(t_{0}\right)\\
\vdots\\
\dot{x}\left(t_{N-1}\right)
\end{array}\right]\approx\mathcal{D}\cdot\text{vec}\left(X\right),
\]
where $\text{vec}\left(\cdot\right)$ is the vectorization operator
that reshapes a matrix into a column vector, and
\[
\mathcal{D}:=D\otimes I_{n_{v}\times n_{p}},
\]
where
\[
D:=\frac{1}{\Delta t}\left[\begin{array}{ccccc}
-1 & 1 &  & \cdots & 0\\
\vdots & -1 & 1 &  & \vdots\\
 &  & \ddots & \ddots\\
0 & \cdots &  & -1 & 1
\end{array}\right],
\]
and $\pi^{N}$ consists of uniformly spaces times samples with
\[
\Delta t=\frac{T}{N}.
\]
Many candidates for $D$ exist, for example higher-order finite differencing schemes \cite{mathews2004numerical} as well as trajectory representations for both uniformly spaced \cite{cichella2018bernstein} and non-uniform time grids \cite{elnagar1995pseudospectral}. These can increase computational accuracy and can be applied to what follows but are outside the scope of this work.

The model parameters, $\theta$, can now be found from the data with
the following constrained optimization problem:
\begin{equation}
\begin{cases}
\underset{X,\theta}{\min} & \left\Vert \text{vec}\left(\delta Y\right)-\text{vec}\left(H\right)\right\Vert _{2}^{2}\\
\\
\text{subject to} & \mathcal{D}\cdot\text{vec}\cdot\left(X\right)-F\left(X;\hat{y},\theta\right)=0,\\
\\
 & lb\leq g\left(X,\theta\right)\leq ub,
\end{cases}\label{eq:single run optimization}
\end{equation}
where the norm squared objective function is justified by the fact
that the measurement noise is assumed Gaussian. In general, the optimization problem above includes a general set of inequality constraints, denoted $g\left(X,\theta\right)$, for specific needs of a pilot modeling task. Common examples are restrictions in the pilot output, $y_p$, such as joystick displacement limits, and constraints on the model parameters to ensure the model remains stable. 

\subsection{Aggregation of Multiple Data Collections}\label{multiple}

Due to variation in human performance, we seek to find the single
model, $\theta$, that best describes all observed data. We assume
that $M$ experiments are conducted and we denote by $X^{j}$, $\delta Y^{j}$,
and $H^{j}$ as the quantities defined above for each $j\in\left\{ 1,2,\ldots,M\right\} $.
The goal target is not assumed constant during each run, with the
specific goal output denoted $\hat{y}^{j}$ for the $j$-th run. The
optimization problem $\left(\ref{eq:single run optimization}\right)$
generalizes to
\begin{equation}
\begin{cases}
\underset{X^{1},\ldots,X^{M},\theta}{\min} & \sum_{j=1}^{M}\left\Vert \text{vec}\left(\delta Y^{j}\right)-\text{vec}\left(H^{j}\right)\right\Vert _{2}^{2}\\
\\
\text{subject to} & \mathcal{D}\cdot\text{vec}\left(X^{1}\right)-F\left(X^{1};\hat{y}^{1},\theta\right)=0,\\
 & \qquad\qquad\qquad\vdots\\
 & \mathcal{D}\cdot\text{vec}\left(X^{M}\right)-F\left(X^{M};\hat{y}^{M},\theta\right)=0,\\
\\
 & lb\leq g\left(X,\theta\right)\leq ub.
\end{cases}\label{eq:multiple run optimization}
\end{equation}

\section{Experiment and Results}

We estimate the pilot model for the following scenario: The pilot is tasked with moving a quadcopter drone, initially at hover, and bringing it to rest at a displaced goal position. The pilot commands a desired reference pitch angle to initiate translational movement towards the goal. The quadcopter is assumed to have an internal stabilizing feedback controller that uses differential throttle (from hover) between the front and back motors to track the reference the pitch angle.

\subsection{Quadcopter Vehicle Model}\label{QuadDyn}
The vehicle chosen for this example is a quadcopter drone, and for illustration purposes we only consider the dynamics associated with the pitch axis. The model is based on what appeared in \cite{bowerfind2025model}. Define the following state vector
\[
x_{v}:=\left[\begin{array}{c}
z\\
\phi\\
v\\
q\\
\omega_{f}\\
\omega_{b}
\end{array}\right]\in\mathbb{R}^6,
\]
where $z$ is the position of the center of mass in the inertial frame, $\phi$ is the pitch angle, $v$ is the (forward) body velocity, $q$ is the body pitch rate, and $\omega_{f},\,\omega_{b}$ are the rotational speeds of the front and back propellors, respectively. The vehicle dynamics are given as
\begin{equation}    
f_{v}\left(x_{v},u_v\right)=\left[\begin{array}{c}
v\\
q\\
\frac{2}{m}\left(T_{f}\left(\omega_f\right)+T_{b}\left(\omega_b\right)\right)\sin\left(\phi\right)\\
\sqrt{2}I_{yy}^{-1}\left(T_{f}\left(\omega_f\right)-T_{b}\left(\omega_b\right)\right)\\
f_{\text{prop}}\left(\omega_{f},e_{f}\right)\\
f_{\text{prop}}\left(\omega_{b},e_{b}\right)
\end{array}\right],\label{eq:quad dynamics}
\end{equation}

where $T_{f}\left(\omega_f\right)$ and $T_{b}\left(\omega_b\right)$ are the thrust from the front and back propellers and where
\begin{equation}
f_{\text{prop}}\left(p,e\right)=-14.48p+0.0222\left(e-1601\right),\label{eq:prop dyn}
\end{equation}
are the propeller dynamics with $e$ being throttle pertubation from the hover equilibrium state. The coefficients of the above propeller model \eqref{eq:prop dyn}, as well as the espressions for thrusts, $T_{f}\left(\omega_f\right)$ and $T_{b}\left(\omega_b\right)$, were obtained empirically from a load cell and rpm data\cite{tyto1580}. The throttle pertubations are calculated as part of a feedback reference
tracking controller with
\[
\left[\begin{array}{c}
e_{f}\\
e_{b}
\end{array}\right]=K\left(\left[\begin{array}{c}
\frac{\pi}{4}u_{v}\\
0
\end{array}\right]-\left[\begin{array}{c}
\phi\\
q
\end{array}\right]\right),
\]
where $K$ was calculated to stabilize the pitch angle, $\phi$,
and was found to be
\[
K=\left[\begin{array}{cc}
-76.37 & -1.02\end{array}\right].
\]

Additional details of the quadcopter model, in particular the coefficients such as mass and moment of inertia, can be found in \cite{bowerfind2025model}.

\begin{figure}\label{fig:sim}
\centering
\includegraphics[width=\linewidth]{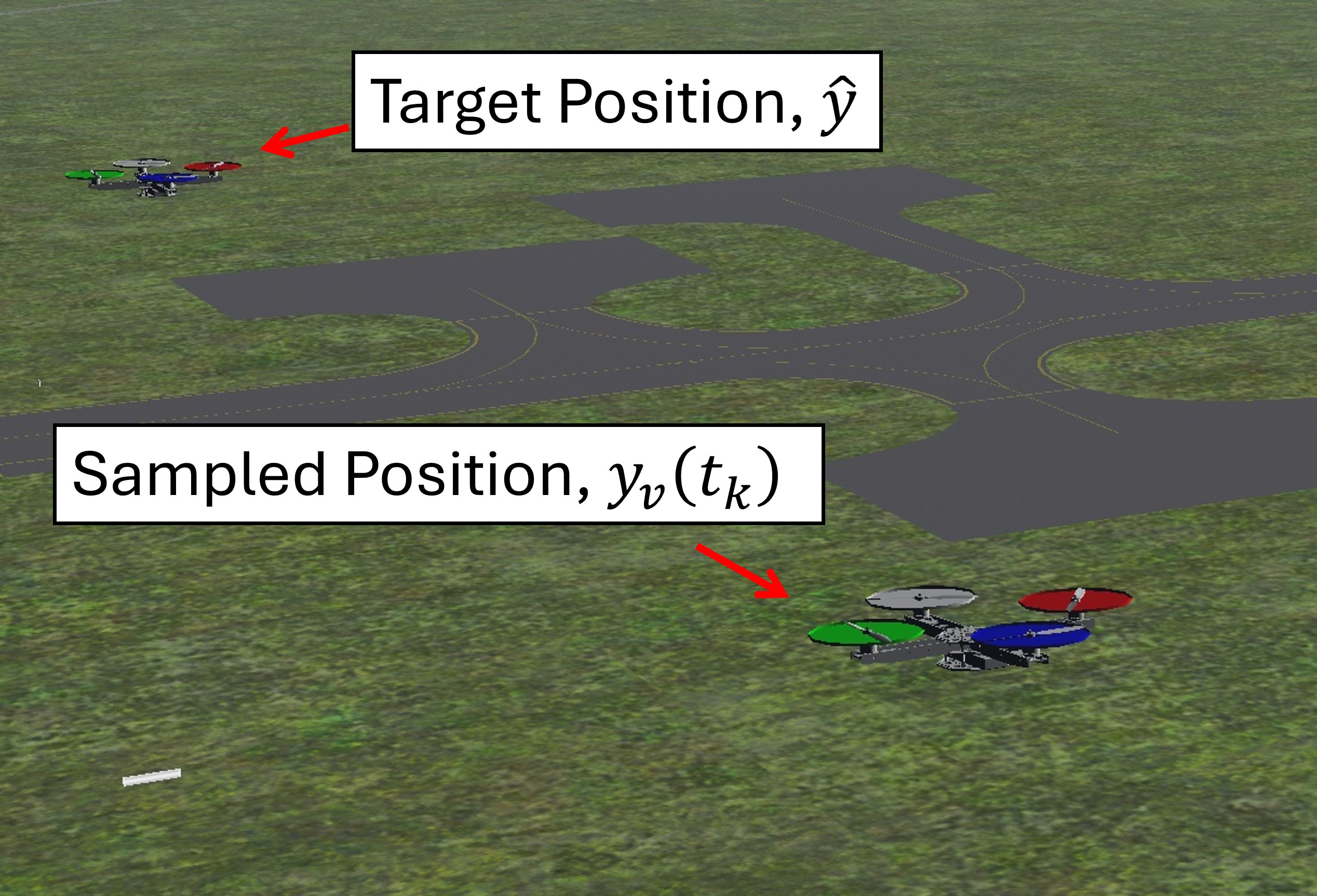}\\
\caption{A screen capture of the simulation environment used for the example in Section \ref{results}. A marker representing the target position is plotted on screen and the pilot, starting from rest, maneuvers the vehicle (in the foreground) to stop directly on top of the target.}\label{fig:sim environment}
\end{figure}

\subsection{Quadcopter Pilot Model}\label{Exp}

We model the pilot with a 3rd-order, single-input, single-output linear system. This system class can be uniquely defined by six scalar parameters giving $\theta\in\mathbb{R}^6$. Several state-space realization exist and without loss of generality we choose the controllable canonical form \cite{hespanha2018linear}:
\begin{align}
f_{p}\left(x_{p},u_{p};\theta\right)= & \left[\begin{array}{ccc}
0 & 1 & 0\\
0 & 0 & 1\\
-\theta_{1} & -\theta_{2} & -\theta_{3}
\end{array}\right]x_{p}\label{eq:linear pilot dynamics}\\
 & +\left[\begin{array}{c}
0\\
0\\
1
\end{array}\right]u_{p},\nonumber
\end{align}
with output
\[
h\left(x_{p};\theta\right)=\left[\begin{array}{ccc}
\theta_{4} & \theta_{5} & \theta_{6}\end{array}\right]x_{p}.
\]
The input for the pilot, $u_p$, is given as $\hat{y}-y_v(t_k)$, i.e. the lateral displacement of the quadcopter relative to the goal position. The output is the commanded pitch angle, normalized to $\left[ -1,1\right]$.

Note that we require the pilot model to be stable. This is achieved when the matrix in \eqref{eq:linear pilot dynamics} is Hurwitz. Since the pilot system is in controllable canonical form, it is a companion matrix, and implies the following characteristic polynomial:
\begin{equation}
    P\left(\lambda\right)=\lambda^3+\theta_3\lambda^2+\theta_2\lambda+\theta_1.\label{eq:charc. polyn}
\end{equation}
We require that the roots of \eqref{eq:charc. polyn} be negative. This is found from the Routh-Hurwitz criterion \cite{routh1877treatise,franklinfeedback} and gives the following conditions on $\theta$:
\begin{equation}    
\begin{cases}
\theta_{1},\theta_{2},\theta_{3}>0,\\
\theta_{3}\theta_{2}-\theta_{1}>0.
\end{cases}\label{eq:routh stability conditions}
\end{equation}  

\subsection{Pilot Model Estimation}

 The vehicle dynamics \eqref{eq:quad dynamics} are implemented in a real-time simulation using Simulink where the position and orientation are displayed graphically using FlightGear \cite{basler2010flightgear}. A joystick is used by the pilot to provide a pitch command. A graphical marker is placed at a distance $\hat{y}$ from the quadcopter in the simulation. This is shown in Figure \ref{fig:sim environment}. The pilot is tasked with maneuvering the quadcopter to the marker. The displacement, $\hat{y}-y_v(t_k)$, is logged at a rate of 20Hz over a time span of 15 seconds for each run.
 
 Three independent runs were recorded for $\hat{y}=10$ meters. Three additional independent runs were recorded with $\hat{y}=20$ meters. A single model, common among all runs, was estimated using \eqref{eq:multiple run optimization}. Additional constraints, $g\left(X,\theta\right)$, were utilized for this experiment, specifically the constraints on $\theta$ given in \eqref{eq:routh stability conditions}. Additionally, the constraint
 \[
 -1\leq h\left(x_{p};\theta\right) \leq 1,
 \]
 was appended to $g\left(X,\theta\right)$. This was added since the joystick used by the pilot is limited to the range $[-1,1]$. The optimization was implemented in Matlab and solved using the NLP code IPOPT \cite{wachter2006implementation}, where the constraint Jacobian and Hessian of Lagrangian \cite{boggs1995sequential} was computed using automatic differentiation by the methods of \cite{andersson2019casadi}.

The resulting pilot model was found to be
\begin{align*}
f_{p}\left(x_{p},u_{p}\right)= & \left[\begin{array}{ccc}
0 & 1 & 0\\
0 & 0 & 1\\
-0.2454 & -1.2477 & -1.0164
\end{array}\right]x_{p}\\
 & +\left[\begin{array}{c}
0\\
0\\
1
\end{array}\right]u_{p},
\end{align*}
with the output equation
\[
y_{p}=\left[\begin{array}{ccc}
-0.0005 & 0.0052 & -0.0575\end{array}\right]x_{p}.
\]

\begin{figure}\label{results}
\centering
\includegraphics[width=\linewidth]{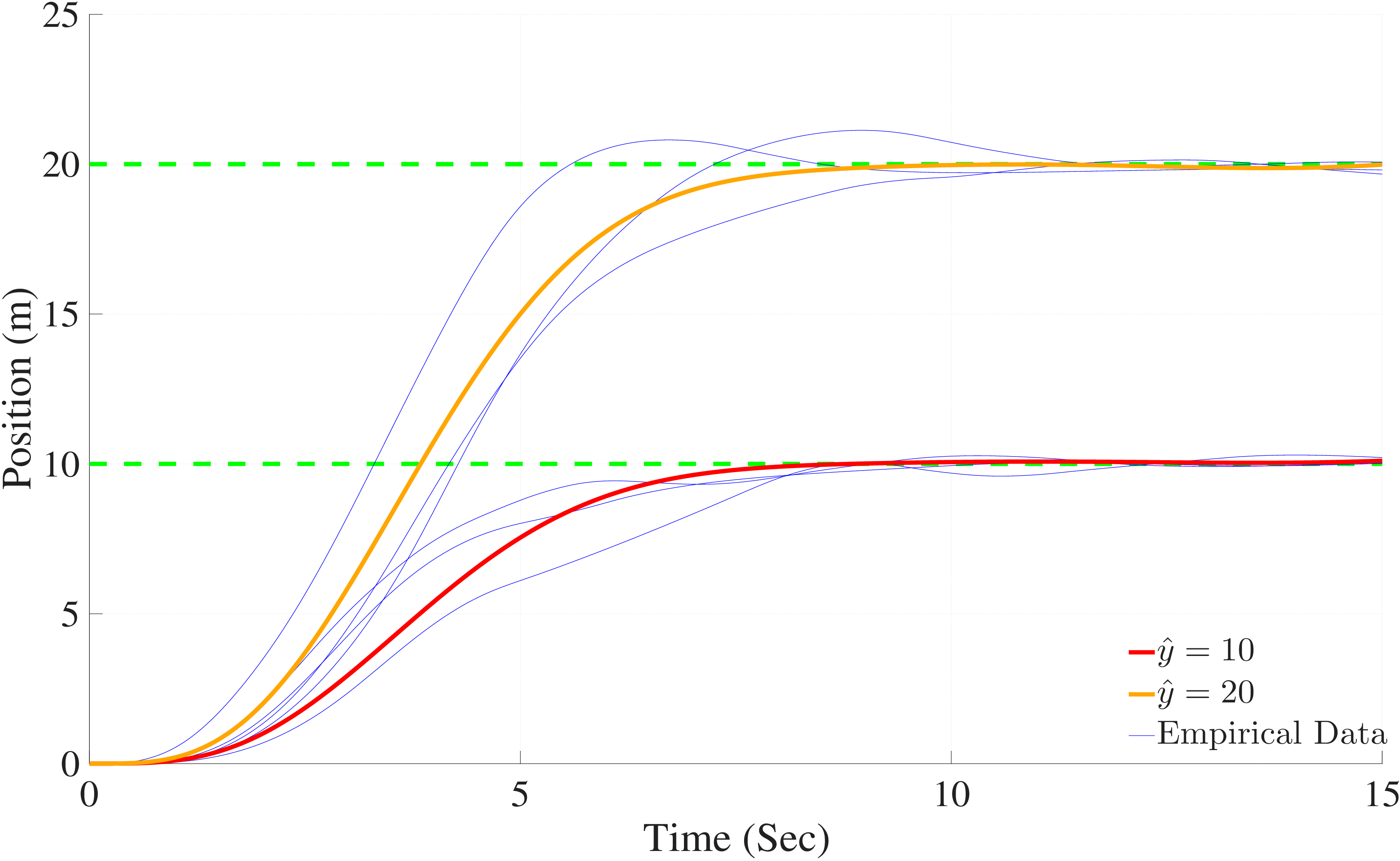}
\caption{The data that was recorded when the pilot was controlling the quadcopter and shown in blue. The green lines show goal positions. The estimated model's response to the same scenarios are shown in red for a 10m goal and shown in yellow for the 20m goal. The same single model was used for both the yellow and green goal states. This shows the quality of the model fit is robust to the natural variation from human controlled vehicles.}
\end{figure}

A plot of the data used to find the model, along with the predicted response of the model itself, is shown in Figure 4.

\section{Conclusion and Future Work}
This paper presents a method for estimating a dynamic model for a human pilot. The method allows for multiple data collections and finds the single model that is common among all experiments. The model parameters are found from a constrained optimization problem which can be solved efficiently using existing NLP algorithms. Future work includes experiments with multiple-input and multiple-output pilot scenarios.

\acknowledgements 
The authors thank Nicoletta Fala, an Assistant Professor of Aerospace Engineering at Auburn University, for helpful suggestions on plotting graphical targets for pilot goal visualization.

This research was supported by the Office of Naval Research under Grant: N00014-24-1-2322.

\bibliographystyle{IEEEtran}
\bibliography{main}

\begin{thebibliography}{10}
\providecommand{\url}[1]{#1}
\csname url@samestyle\endcsname
\providecommand{\newblock}{\relax}
\providecommand{\bibinfo}[2]{#2}
\providecommand{\BIBentrySTDinterwordspacing}{\spaceskip=0pt\relax}
\providecommand{\BIBentryALTinterwordstretchfactor}{4}
\providecommand{\BIBentryALTinterwordspacing}{\spaceskip=\fontdimen2\font plus
\BIBentryALTinterwordstretchfactor\fontdimen3\font minus \fontdimen4\font\relax}
\providecommand{\BIBforeignlanguage}[2]{{%
\expandafter\ifx\csname l@#1\endcsname\relax
\typeout{** WARNING: IEEEtran.bst: No hyphenation pattern has been}%
\typeout{** loaded for the language `#1'. Using the pattern for}%
\typeout{** the default language instead.}%
\else
\language=\csname l@#1\endcsname
\fi
#2}}
\providecommand{\BIBdecl}{\relax}
\BIBdecl

\bibitem{kirchner2020reachability}
M.~R. Kirchner, E.~Ball, J.~Hoffler, and D.~Gaublomme, ``Reachability as a unifying framework for computing helicopter safe operating conditions and autonomous emergency landing,'' \emph{IFAC-PapersOnLine}, vol.~53, no.~2, pp. 9282--9287, 2020.

\bibitem{tustin1947nature}
A.~Tustin, ``The nature of the operator's response in manual control, and its implications for controller design,'' \emph{Journal of the Institution of Electrical Engineers-Part IIA: Automatic Regulators and Servo Mechanisms}, vol.~94, no.~2, pp. 190--206, 1947.

\bibitem{mcruer1974mathematical}
D.~T. McRuer and E.~S. Krendel, ``Mathematical models of human pilot behavior,'' Tech. Rep., 1974.

\bibitem{roncolini2023virtual}
F.~Roncolini, G.~Quaranta \emph{et~al.}, ``Virtual pilot: {A} review of the human pilot’s mathematical modeling techniques,'' in \emph{Aerospace Europe Conference 2023-Joint 10th EUCASS-9th CEAS Conference}, 2023, pp. 1--14.

\bibitem{jirgl2015dynamic}
M.~Jirgl, M.~Havlikova, and Z.~Bradac, ``The dynamic pilot behavioral models,'' \emph{Procedia Engineering}, vol. 100, pp. 1192--1197, 2015.

\bibitem{mathews2004numerical}
J.~Mathews, \emph{Numerical Methods Using {Matlab}}, 3rd~ed.\hskip 1em plus 0.5em minus 0.4em\relax Prentice Hall, 1999.

\bibitem{cichella2018bernstein}
V.~Cichella, I.~Kaminer, C.~Walton, N.~Hovakimyan, and A.~Pascoal, ``Bernstein approximation of optimal control problems,'' \emph{arXiv preprint arXiv:1812.06132}, 2018.

\bibitem{elnagar1995pseudospectral}
G.~Elnagar, M.~A. Kazemi, and M.~Razzaghi, ``The pseudospectral {Legendre} method for discretizing optimal control problems,'' \emph{IEEE Transactions on Automatic Control}, vol.~40, no.~10, pp. 1793--1796, 1995.

\bibitem{bowerfind2025model}
S.~R. Bowerfind, M.~R. Kirchner, G.~A. Hewer, D.~R. Robinson, P.~Chen, A.~Farahmandi, and K.~Estabridis, ``A model-free data-driven algorithm for continuous-time control,'' in \emph{2025 IEEE Aerospace Conference}, 2025, pp. 1--10.

\bibitem{tyto1580}
\emph{Series {1580/1585} User Manual}, V2.0~ed., Tyto Robotics, 2021.

\bibitem{hespanha2018linear}
J.~P. Hespanha, \emph{Linear Systems Theory}.\hskip 1em plus 0.5em minus 0.4em\relax Princeton University Press, 2018.

\bibitem{routh1877treatise}
E.~J. Routh, \emph{A treatise on the stability of a given state of motion, particularly steady motion}.\hskip 1em plus 0.5em minus 0.4em\relax Macmillan and Company, 1877.

\bibitem{franklinfeedback}
G.~F. Franklin, J.~D. Powell, and A.~Emami-Naeini, \emph{Feedback Control of Dynamic Systems}, 5th~ed.\hskip 1em plus 0.5em minus 0.4em\relax Pearson Prentice Hall, 2006.

\bibitem{basler2010flightgear}
M.~Basler, M.~Spott, S.~Buchanan, J.~Berndt, B.~Buckel, C.~Moore, C.~Olson, D.~Perry, M.~Selig, D.~Walisser \emph{et~al.}, ``The {FlightGear} manual,'' \emph{The FlightGear Manual}, 2010.

\bibitem{wachter2006implementation}
A.~W{\"a}chter and L.~T. Biegler, ``On the implementation of an interior-point filter line-search algorithm for large-scale nonlinear programming,'' \emph{Mathematical Programming}, vol. 106, pp. 25--57, 2006.

\bibitem{boggs1995sequential}
P.~T. Boggs and J.~W. Tolle, ``Sequential quadratic programming,'' \emph{Acta Numerica}, vol.~4, pp. 1--51, 1995.

\bibitem{andersson2019casadi}
J.~A. Andersson, J.~Gillis, G.~Horn, J.~B. Rawlings, and M.~Diehl, ``{CasADi}: {A} software framework for nonlinear optimization and optimal control,'' \emph{Mathematical Programming Computation}, vol.~11, pp. 1--36, 2019.

\end{thebibliography}




\thebiography
\begin{biographywithpic}
{Harrison M. Bonner}{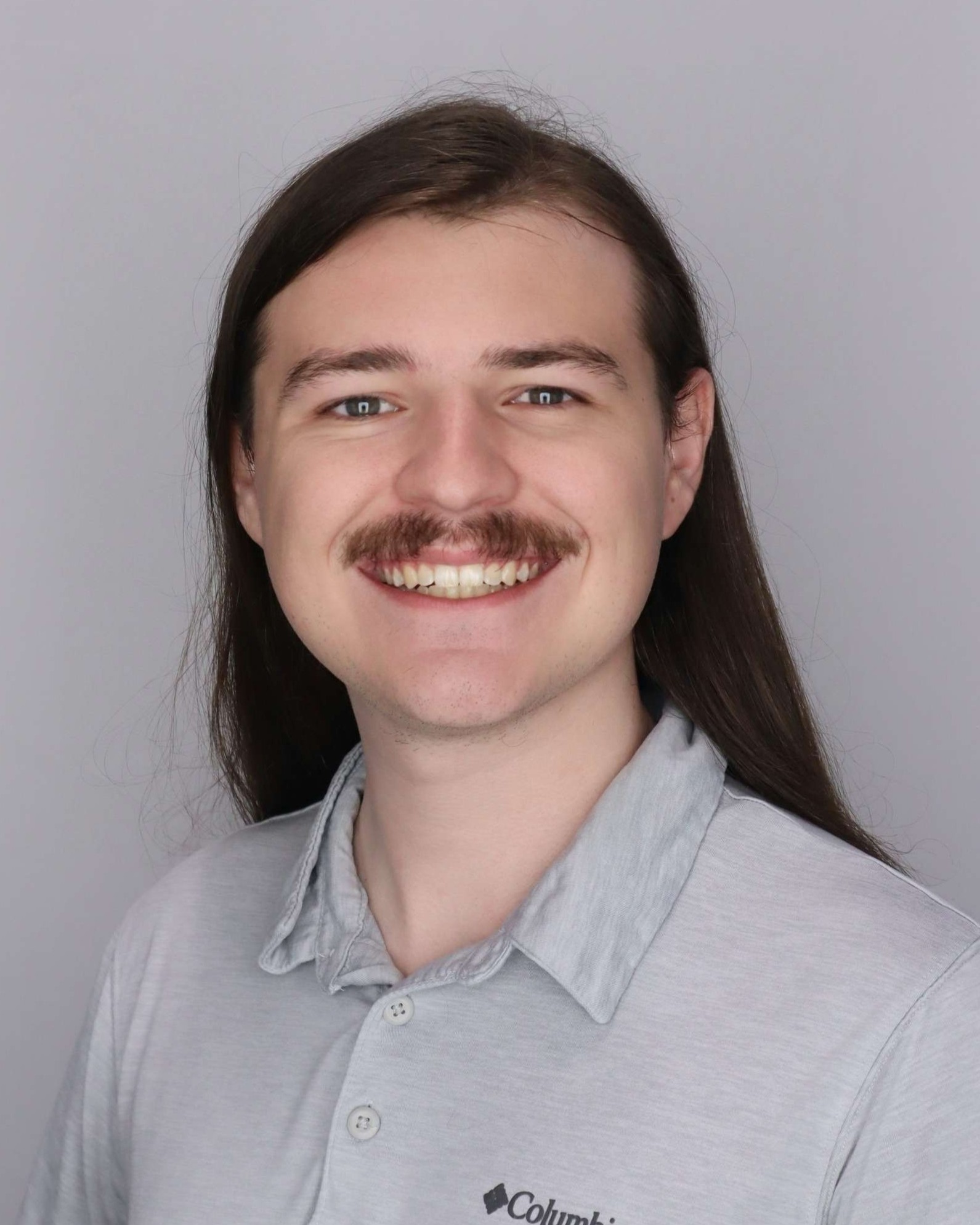} is
an undergraduate research assistant in Auburn University's Autonomous Systems Lab and currently pursuing a bachelor’s degree in computer engineering. His research interests span robotics, optimal control, and guidance and navigation systems. Harrison has received numerous accolades at Auburn University including the Presidential Scholarship (2023-2025), the Drummond Company Honors Endowed Scholarship (2024-2025), the Pumphrey Outstanding Pre-Engineering Student Award (2024), as well as being selected for the Dean's List (2023-2025) and Honors College Director’s List (2025).
 
\end{biographywithpic} 

\begin{biographywithpic}
{Matthew R. Kirchner}{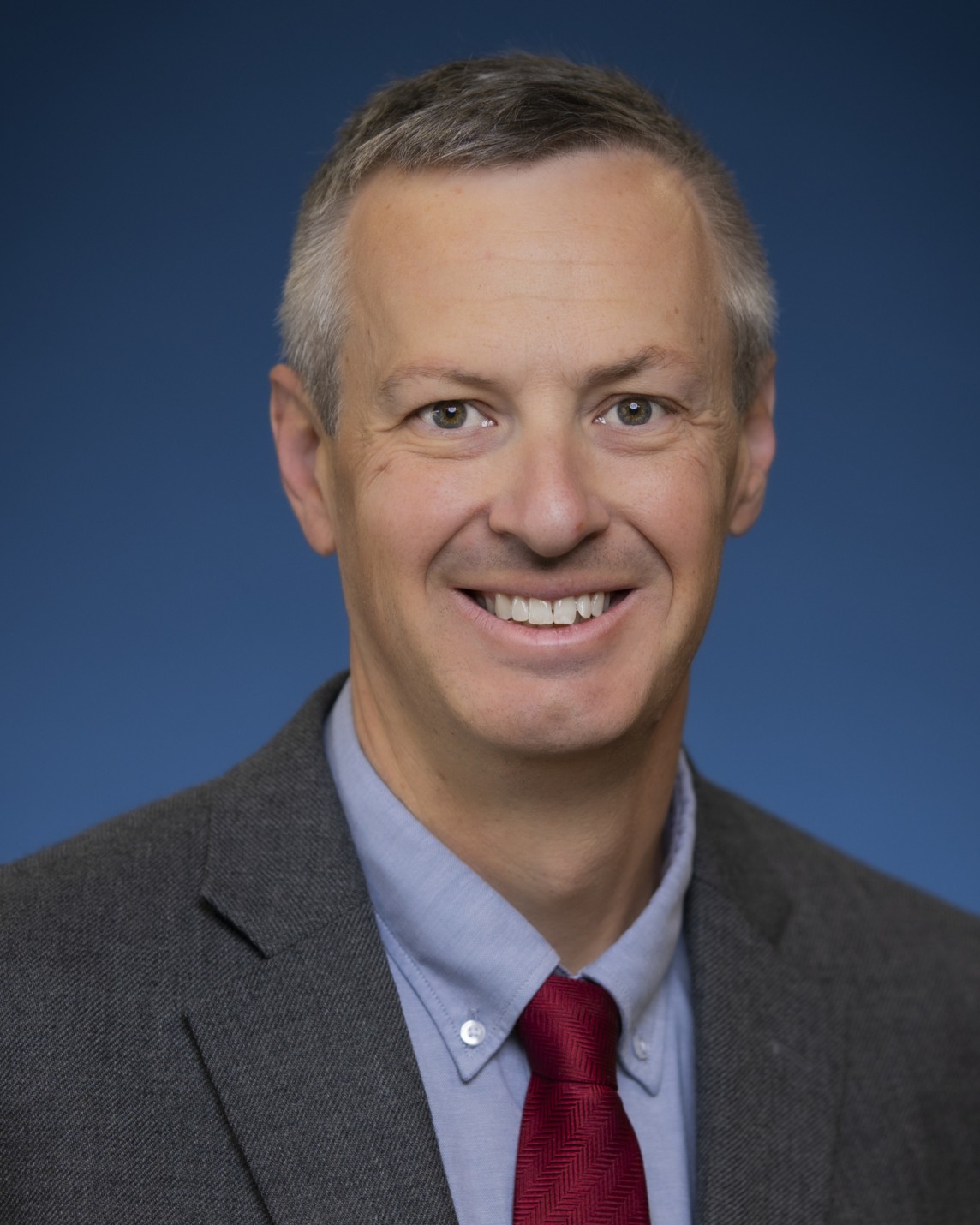}
is a Godbold Endowed Assistant Professor of Electrical and Computer Engineering at Auburn University. He received his B.S. in mechanical engineering from Washington State University in 2007, a M.S. in electrical engineering from the University of Colorado at Boulder in 2013, and a Ph.D. in electrical engineering from the University of California, Santa Barbara in 2023. He spent over 16 years at the Naval Air Warfare Center Weapons Division, China Lake, first joining the Navigation and Weapons Concepts Develop Branch in 2007 as a staff engineer. In 2012 he transferred into the Physics and Computational Sciences Division within the Research and Intelligence Department, where he served as a senior research scientist. His research interests include level set methods for optimal control, differential games, and reachability; multi-vehicle robotics; nonparametric signal and image processing; and navigation and flight control. He was the recipient of a Naval Air Warfare Center Weapons Division Graduate Academic Fellowship from 2010 to 2012; in 2011 was named a Paul Harris Fellow by Rotary International and in 2021 was awarded a Robertson Fellowship from the University of California in recognition of an outstanding academic record.
\end{biographywithpic}

\balance

\end{document}